\documentclass[%
reprint,
superscriptaddress,
 amsmath,amssymb,
 aps,
 prc,
longbibliography]{revtex4-1}
\usepackage{comment}
\usepackage{graphicx}
\usepackage{mathtools}
\usepackage{subfigure}
\usepackage{braket}
\usepackage{dcolumn}
\usepackage{bm}
\usepackage{epstopdf}
\usepackage{xcolor}
\usepackage{tikz}
\usetikzlibrary{external}
\usepackage{hyperref} 
\usepackage{appendix} 
\usepackage{float} 

\usepackage{ulem}

\begin{document}

\title{An alternative representation of multichannel Rydberg spectra: \\a modified K-matrix and Lu-Fano plot, applied to manganese spectroscopy}

\author{Justin D. Piel}
\affiliation{Department of Physics and Astronomy, Purdue University, West Lafayette, Indiana 47907 USA}
\author{Chris H. Greene}
\affiliation{Department of Physics and Astronomy, Purdue University, West Lafayette, Indiana 47907 USA}
\affiliation{Purdue Quantum Science and Engineering Institute, Purdue University, West Lafayette, Indiana 47907 USA}

\date{\today}

\begin{abstract}
The well-known graphical representation called the Lu-Fano plot was originally developed for multichannel Rydberg spectroscopy, especially in quantum defect theory. The present study shows some of the limitations of this traditional Lu-Fano plot that are desirable to improve on, when there are closely split ionization thresholds as in many current generation quantum information applications involving hyperfine-split thresholds, or when there are more than two ionization threshold energies. The modified representation introduced here is especially simplifying in the situation where one is exploring the bound states lying very far below those closely split thresholds.  Moreover, it overcomes one limitation, namely that in contrast to the traditional Lu-Fano plot, the modified representation developed here can be utilized for problems where there more than two ionization thresholds.  An example application to Rydberg series of the manganese atom illuminates its use in a practical problem.
\end{abstract}

\pacs{Valid PACS appear here}
\maketitle

\section{Introduction}
In early studies of multichannel Rydberg series, especially those by Fano, Lu, Lee, and others,\cite{Fano_1970,Lu_Fano_1970,Lu_1971,Lee_Lu_1973,BrownGinter1975,Armstrong_Esherick_Wynne_1977,mullins1985channel} a common situation was explored in detail, where there are some finite number of Rydberg series that converge to just two ionization thresholds.  The Lu-Fano plot, as was introduced in \cite{Lu_Fano_1970}, treated the $J^\pi=1^-$ symmetry of atomic xenon, which has five Rydberg series converging to two spin-orbit-split $(^2P_{3/2},^2P_{1/2})$ thresholds separated by approximately 1.3 eV.  The Lu-Fano plot depicts the quantum defect (modulo 1) of Rydberg energy levels relative to the lower energy threshold versus the effective quantum number relative to the upper threshold.  This plot is approximately periodic in each axis and is sometimes plotted modulo 1 for both axes, in the periodic version of such plots.  One sees at a glance that there are frequently avoided crossings and interesting regions that signal where channel interactions are particularly strong or weak.
Atomic units are used throughout this paper unless specified otherwise.


\section{Theoretical Development}
Recall that the main conceptual idea of multichannel quantum defect theory (MQDT) is to postpone application of the long-range boundary condition (vanishing at $r\rightarrow \infty$ in all energetically closed channels) since that channel closing step can be done trivially in the final stage of the calculation.  The benefit of this postponement in the theoretical description is that the Rydberg electron wavefunctions display no back-reflected de Broglie waves from the outermost radial turning point that are responsible for highly energy-dependent interferences; consequently the physics is controlled by a scattering matrix $S$ or a reaction matrix $K$, which are generally smooth in energy.  These matrices used in multichannel quantum defect theory (MQDT)\cite{Seaton_1966,FanoJOSA} are sometimes referred to as ``unphysical'' because they include channels treated temporarily as ``open'', even though their ionization thresholds $E_i$ are energetically closed at an energy $E$ of interest, i.e., having $E<E_i$.  Such a closed channel should never be present in a physical scattering matrix, of course, because one could never fire an electron from infinity to initiate a collision in such a channel. 
The ability to turn a table of energy level values into a graphical grouping of those energies along mathematically-defined curves has been a long-standing reason why Lu-Fano plots have become a staple for presenting multichannel Rydberg spectra with interpretive power.  
The rotated $K$-matrix ($\tilde{K}$) and modified Lu-Fano (MLF) plot introduced below extend that interpretive power into different regimes where the original version has some severe limitations. As can be seen below, the construction of $\tilde{K}$ has some similarities but also major differences from the phase-shifted variants of MQDT described in Refs. \cite{lee2009decoupling,Lee_2017,Giusti-Suzor_Fano_1984,Cooke_Cromer_1985}. 
\vspace{-10pt}
\subsection{Review of needed Coulomb radial solution properties}
The equations of MQDT are expressed most simply in terms of the ``energy-normalized'' base pair of spherical Coulomb radial wavefunctions, typically denoted $(f,g)$ in Refs.\cite{Greene_Fano_Strinati_1979,Greene_Rau_Fano_1982,Greene_Rau_Fano_1984, OrangeReview} which correspond to the solutions denoted $(s,-c)$ by Seaton \cite{Seaton_RPP_1983}.  Here $f$ or $s$ represents the solution regular at the origin, $r=0$.   The most important property of these solutions for a discussion of multichannel bound states is their asymptotic forms at negative channel energies, namely:
\begin{eqnarray}
    f_i(r)  \xrightarrow[r\to\infty]{} C_i r^{-\nu_i}e^{r/\nu_i} \sin{\beta_i} + O(r^{\nu_i} e^{-r/\nu_i})  \\  
    g_i(r)  \xrightarrow[r\to\infty]{} -C_i r^{-\nu_i}e^{r/\nu_i} \cos{\beta_i} + O(r^{\nu_i} e^{-r/\nu_i}),  \label{fgAsymptotic}
\end{eqnarray}
where $C_i\equiv C(\nu_i,\ell_i)$ is a constant that can be found in the aforementioned references on Coulomb functions and is largely irrelevant in the following.  The channel energy (in a.u.) for Rydberg channel $i$ is denoted $\epsilon_i \equiv -\frac{1}{2{\nu_i}^2}$, and the negative energy phase parameter is defined as $\beta_i \equiv \pi(\nu_i - \ell_i)$ for channel $i$ having orbital angular momentum $\ell_i$.  The constant $\nu_i$ is referred to as the {\it effective quantum number} in channel $i$, and is a recast energy variable that becomes infinite at the ionization threshold of the $i$-th channel where $\epsilon_i \rightarrow 0$.  Note that the Coulomb function properties discussed throughout the present article are for an electron that experiences a Coulomb potential with unit positive charge asymptotically, though the concepts immediately apply to different ionic charge states $Z>1$ with minor modifications.

Multichannel quantum defect theory temporarily characterizes solutions in the Coulomb zone outside some radius $r_0$ in terms of these radial solutions $(f_i,g_i)$, even though they are exponentially divergent at $r\rightarrow \infty$; the vanishing at $\infty$ is imposed later analytically in the so-called ``closed-channel elimination'' step of MQDT, which simply relies on the fact that the linear combination in each channel that is regular at $\infty$ is simply 
\begin{equation}
    W_i(r) = f_i(r) \cos{\beta_i}+g_i(r) \sin{\beta_i}.\label{Whittaker0}
\end{equation}
This is referred to as the ``energy-normalized'' Whittaker function.\cite{OrangeReview}

A common, standardized way to represent the Rydberg electron solutions that arise in a multichannel situation with multiple ionization thresholds $E_i$ with $i=1, 2,...N$ is in terms of a real, symmetric $N \times N$ reaction matrix $K_{i,i'}$.  Equivalently, the Fano school often preferred to discuss the eigenvalues ($\tan{\pi \mu_\alpha}$ and corresponding eigenvectors $U_{i \alpha}$, denoting the $\mu_\alpha$ as eigen quantum defects.\cite{FanoJOSA}  Prior to imposition of decaying boundary conditions, the $i'$-th linearly independent solution in the purely Coulombic region $r>r_0$ is written:
\begin{equation}
    \psi_{i'}=  {\cal A}\frac{1}{r} \sum_i \Phi_i(\Omega) (f_i(r) \delta_{ii'}-g_i(r) K_{i,i'}). \label{KmatrixSolutions}
\end{equation}
Here $\cal A$ is an antisymmetrization operator, and $r$ denotes the radial coordinate of the outermost atomic or molecular electron in the system.  The channel function $\Phi_i$ is an eigenfunction of all remaining degrees of freedom ($\Omega$) besides $r$ that fully characterize one ionization channel, including typically the ionic energy eigenstate associated with channel $i$, times the angular and spin eigenstates of the outermost electron, with their relevant angular momentum coupling factors.

While MQDT routinely handles all energy ranges, such as above and below thresholds, in bound, autoionizing, and open continuum energy ranges, the present study focuses on the energy range far below multiple closely spaced ionization thresholds $E_i$.  Each wavefunction in Eq. \ref{KmatrixSolutions} in the energy range where the energy $E$ lies below all of the threshold energies $E_i$ diverges at infinity, and the actual bound energy levels of the system occur only when a linear combination of the independent solutions vanishes at $r\rightarrow \infty$.  Representing a desired energy eigenstate as $\Psi = \sum_{i'} \psi_{i'} A_{i'}$, and inserting Eqs.\ref{fgAsymptotic}, gives a homogeneous linear system whose nontrivial determinantal roots determine the physical bound state energies, i.e.
\begin{equation}
\label{eq:linearsystem1}
    (\underline{K} + \tan\underline{\beta})\overrightarrow{A}=0,
\end{equation}
where the notation $\underline{K}$ indicates that $K$ is a matrix.
Typically the $K$-matrix is a smooth function of energy, and over small energy ranges relevant to Rydberg state physics it can even be approximated as an energy-independent constant matrix.  The matrix $\underline{\beta}$ is diagonal and consists of the Coulomb phase parameters in each channel, as defined above in terms of the effective quantum numbers, which of course satisfy the  additional constraints, namely 
\begin{equation}
\label{eq:Rydbergform}
    E = E_i - \frac{1}{2 \nu_i^2},
\end{equation}
for all closed channels $i$.  In the form of Eq. \ref{eq:linearsystem1}, the bound state quantization is not very convenient in practice because of the periodic divergences of $\tan{\pi(\nu_i-\ell_i)}=\tan{\pi \nu_i}$.  This is ameliorated by left multiplying Eq. \ref{eq:linearsystem1} by $\cos{\pi(\nu_i-\ell_i)}$, giving the more stable form for root finding:
\begin{equation}\label{eq:linearsystem2}
     (\cos{ \underline{\beta}}(E)\underline{K} + \sin{ \underline{\beta}}(E))\overrightarrow{A}=0. 
\end{equation}
When an energy-dependent $K$-matrix is utilized, as is necessary in some problems, there could also be occasional divergent poles of $K(E)$ at certain energies.  Whenever that is an issue, one can either recast the linear system in terms of $K^{-1}$ or else solve this linear system in the eigenrepresentation of $K$.  This recasts the linear system in terms of the eigenvalues of $K$, written in terms of eigen quantum defects as $\tan{\pi\mu_\alpha}$, and their corresponding eigenvectors $U_{i\alpha}$, as is discussed below.
Much of the present article will make use of the flexibility in choosing different options for the base pair $(f,g)$ of linearly-independent radial solutions to the Coulomb plus centrifugal Schr\"odinger equation.

\subsection{Overview of the Traditional Lu-Fano Plot}

For starters, there is both an experimental and a theoretical perspective on creating a Lu-Fano-style plot of interacting Rydberg series.  We first present the way an experimentalist can approach it, armed only with the measured energy levels $E^{(n)}$ and ionization threshold energies $E_i$, $i=1...N_{\rm thresh}$ that are assumed to be known.  Of course if there is only a single ionization threshold $E_1$, one simply computes the effective quantum number $\nu_1^{(n)} \equiv [(E_1-E^{(n)})/Ry]^{-1/2}$ relative to that threshold, and then the quantum defect (mod 1) is $\mu_1^{(n)} = -\nu_1^{(n)}$ {\rm (mod 1)}. $Ry$ is the Rydberg constant for the system, with the appropriate electron-ion reduced mass. That single threshold case would produce the most trivial type of Lu-Fano plot, namely a plot of points with $\mu_1^{(n)}$ on the $y$-axis and $\nu_1^{(n)}$ on the $x$-axis.  

More interesting and central to the ideas of Lu and Fano, however, is a situation where there are two ionization thresholds, with possibly multiple channels attached to each such threshold.  For this case, one still computes the  quantum defect $\mu_1^{(n)}$ for the $n$-th level for the $y$-axis, but the $x$-axis for that level is instead the effective quantum number relative to the second threshold, namely $\nu_2^{(n)} \equiv [(E_2-E^{(n)})/Ry]^{-1/2}$.  This plot of a set of measured points already begins to group similar levels together along a curve that one can begin to imagine.  But the continuous mathematical curve itself that goes through all of those points is not yet known, until the reaction matrix $K$ has been determined, or equivalently, the eigen defects $\mu_\alpha$ and eigenvector matrix $U_{i \alpha}$. That information has sometimes been extracted by fitting to the set of points, but it is the aim of theory to predict that $K$-matrix which yields the curve that goes through every experimental point on the graph.  

There are several ways in which the $K$-matrix can be determined, for instance one can directly solve the Schr\"odinger (or Dirac) equation \cite{OrangeReview} to obtain a theoretical $K$-matrix, or an alternative strategy used by spectroscopists is to empirically fit the experimental levels in the MQDT framework, in some cases in conjunction with a frame transformations. The latter is used throughout this article and a few examples of such fits can be found in Refs. \cite{BrownGinter1975, AymarPhysRept, Armstrong_Esherick_Wynne_1977, GallagherYb, Peper_Li_Knapp_Bileska_Ma_Liu_Peng_Zhang_Horvath_Burgers_2025}.

\begin{figure}
    \centering
    \includegraphics[width=0.8\linewidth]{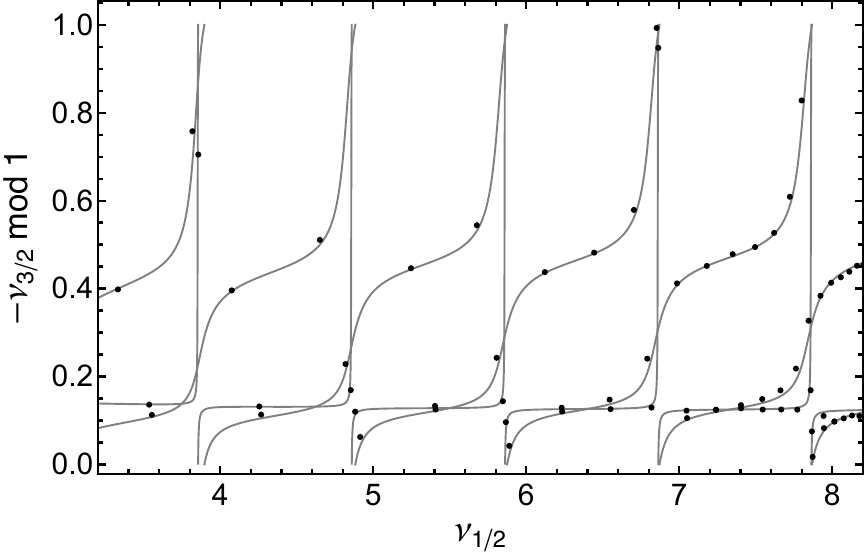}
    \caption{Traditional Lu-Fano plot of the $J^\pi=1^-$ symmetry states of Ar showing the quantum defect -$\nu_{3/2}$ mod 1 vs $\nu_{1/2}$. Black circles are experimental level positions taken from \cite{NIST_ASD}. The $(^2P^o_{3/2})8d\ ^2[\frac{3}{2}]^o_1$ level given in the NIST table as 125736.9 cm$^{-1}$ is a typo and should actually be 125376.9 cm$^{-1}$ \cite{Yoshino_1970}. The MQDT parameters were taken from \cite{Lee_Lu_1973}.}
    \label{fig:ArgonLFplot}
\end{figure}

\begin{figure}
    \centering
    \includegraphics[width=0.7\linewidth]{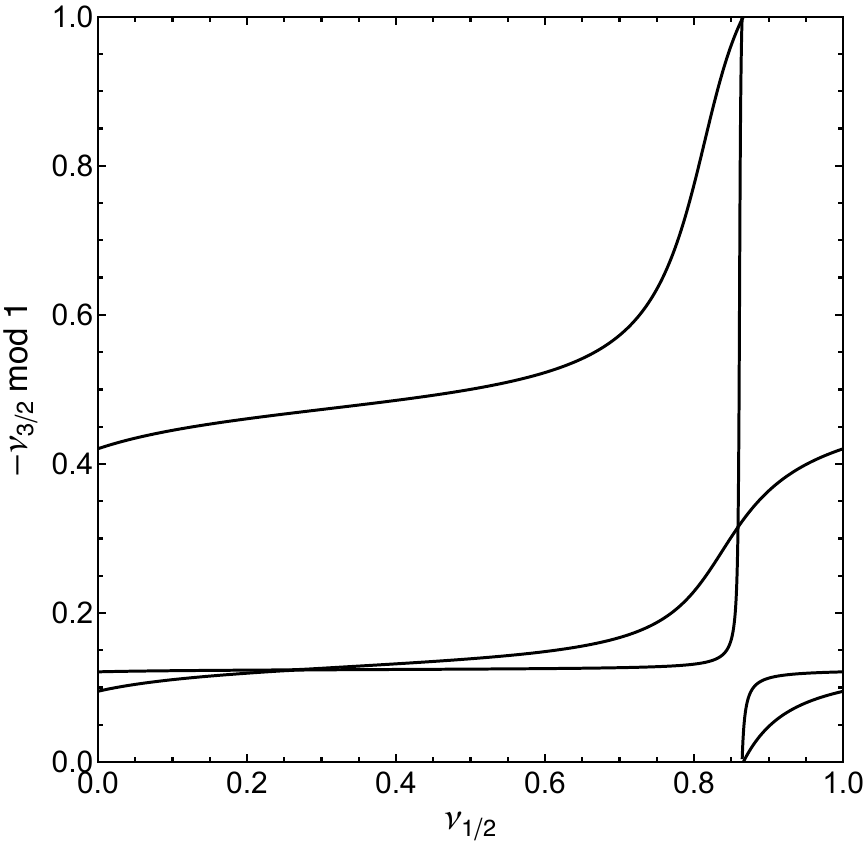}
    \caption{Periodic Lu-Fano plot of Ar using energy independent parameters showing -$\nu_{3/2}$ mod 1 vs $\nu_{1/2}$ mod 1. MQDT parameters were obtained from\cite{Lee_Lu_1973}.}
    \label{fig:ArgonLFplotperiodic}
\end{figure}

Another key theoretical concept that is frequently used in finding the $K$-matrix that yields the smooth curves on a Lu-Fano plot is the frame transformation theory, which makes use of the fact that the short-range part of the Rydberg electron motion often has approximately good quantum numbers that are incompatible with the long-range good quantum numbers.  For the Ar spectrum examples in Figs. \ref{fig:ArgonLFplot} and \ref{fig:ArgonLFplotperiodic}, the short-range approximately good quantum numbers are the spin $S$ and orbital $L$ angular momenta, whereas the long-range quantum numbers are more appropriately chosen to be $jj$ or $jK$ coupled.  Another example of a hyperfine frame transformation is discussed in the following sections.

The style of plot introduced by Lu and Fano \cite{Lu_Fano_1970} in 1970 provides a simple way to visualize the frequently strong interaction among Rydberg series converging to two distinct ionization thresholds.  Such plots are often utilized by both experimentalists and theorists to understand the nature of interacting Rydberg series, but they have been limited to date to analyze two-threshold atomic or molecular systems, aside from rare exceptions described below (e.g. Ref. \cite{Armstrong_Esherick_Wynne_1977}) that considered Rydberg series attached to three or more ionization thresholds.  In order to understand some of the strengths and weaknesses of Lu-Fano-style plots, we begin by describing how they are constructed, which have two main variations.  

The first type of the traditional Lu-Fano plot is for a system with two ionization thresholds, located at energies $E_1 < E_2$.  If there were no channel interactions, i.e. if the $K$-matrix with $N_1$ Rydberg series converging to threshold $E_1$ and $N_2$ series converging to threshold $E_2$ would be diagonal and the diagonal elements would represent quantum defects in every channel, $\mu_i =\frac{1}{\pi} \arctan{K_{ii}}$.  This scenario produces two uncoupled sets of Rydberg series, $N_1$ of which would give energies in a.u. converging to threshold $E_1$, and $N_2$ of which would give energies in a.u. converging to threshold $E_2$, as in the following equations:    
\begin{eqnarray}
    E^{(n)}=E_1-\frac{1}{2(n_1-\mu_i)^2}, i \in {1,....,N_1} \\
    E^{(n)}=E_2- \frac{1}{2(n_2-\mu_i)^2}, i \in {N_1+1,...,N_1+N_2}. 
\end{eqnarray}
These formulas assume that the channels $\{ i \}$ have been ordered with increasing (or equal) threshold energies $E_1 \leq E_2$.

The more interesting and more typical situation involves a non-diagonal $K$-matrix that describes nontrivial channel interactions and level repulsion among the $N=N_1+N_2$ channels.  Given a $K$-matrix, obtained either through theoretical means or empirical fits, 
and the threshold energies, one can find various scattering, photoabsorption, and spectral observables: eg. the allowed energies, for bound states and resonances, in at least two equivalent ways. {\it  Method  (1)} involves searching directly for roots of the determinantal equation, e.g. in the form of Eq. \ref{eq:linearsystem1} or \ref{eq:linearsystem2}, which can become a complicated task of nonlinear root-finding.  {\it  Method (2)
} applies the finiteness boundary condition at $r \rightarrow \infty$ in two separate steps.  The first step performs that channel closing {\it only} for the set of the channels attached to the {\it upper threshold} $E_2$.  This can be viewed as initially treating all channels attached to the lower threshold as though they are energetically open.  The channel elimination can be performed by partitioning the full $K$-matrix into an ``open'' subset (those attached to threshold $E_1$ are treated temporarily as ``open'' when solving the MQDT equations) and a closed subset.  Enforcing exponential decay in all of the upper channels of the closed subset is accomplished by the following algebra that defines an energy-dependent reduced $N_1 \times N_1$ matrix $K^{\rm red}(E)$:
\begin{equation}
  K^{\rm red} = K^{\rm oo}-K^{\rm oc}(K^{\rm cc}+\tan{\pi \nu_c})^{-1}K^{\rm co},  \label{eq:channelelim}
\end{equation}
where $\nu_c(E)$ is a diagonal matrix of the effective quantum numbers in the closed subset channels attached to ionization threshold $E_2$ (and in any closed channel thresholds even higher than $E_2$ if there are more than two thresholds in the system).
The eigenvalues of this reduced dimension reaction matrix $K^{\rm red}(E)$ are now computed and are denoted, in the notation of Lee and Lu \cite{Lee_Lu_1973}, as $\tan{\pi \tau_\rho(E)}$. Note that these eigen defects $\tau_\rho(E)$ with the independent energy variable $E$ replaced by $\nu_2$ form the continuous curves (continuous aside from branch changes) shown in Figs. \ref{fig:ArgonLFplot} and \ref{fig:ArgonLFplotperiodic}, which pass through every experimental energy level on the Lu-Fano plot.  For states lying on a portion of the curve that is mostly horizontal, they belong in zeroth order to the Rydberg series converging to $E_1$, whereas states lying on mostly vertical portions of the continuous curves belong mainly to series converging to $E_2$.

The final step of eliminating exponential growth in the channels attached to threshold $E_1$, which is the point where the mathematics finally determines the true bound state energy eigenvalues of the system, is now  achieved by demanding that $det[K^{\rm red}(E) + \tan{\beta_1} \mathbb{I}]=0$;  in terms of the eigenphaseshifts of $K^{\rm red}$, this bound state condition is represented more simply as the condition: $\tau_\rho(E)+ \nu_1 = n$ where $n$ is an integer. But even before this final step of channel elimination for the lower threshold components, the $\tau_\rho(E)$ are already smooth curves (aside from trivial branch changes) that show the $N_1$ energy-dependent quantum defects in the Rydberg series converging to threshold $E_1$, as affected by interactions with the channels converging to the upper threshold $E_2$.  Fig.\ref{fig:ArgonLFplot} shows an example of a traditional Lu-Fano plot, which plots these quantum defects $\tau_{\rho}(E)$ as a function of the energy variable, usually chosen to be $\nu_2$ on the $x-axis$ of the plot.  A second variation of the traditional Lu-Fano plot, called the {\it periodic version}, is shown for the same system in Fig.\ref{fig:ArgonLFplotperiodic}; where the only difference is that the $\nu_2$-axis is now plotted modulo 1.  This takes advantage of the fact that there are only two trigonometric functions in the MQDT channel elimination step, namely $\tan{\pi \nu_1}$ and $\tan{\pi \nu_2}$, implying that in the limit of an energy-independent $K$-matrix, the system would be perfectly periodic and repeating.  The result is that the entire multichannel Rydberg series, for any atomic or molecular system with only two relevant ionization thresholds, collapses into simple curves in this plot of $\tau_\rho$(mod 1) versus $\nu_2$ (mod 1), as the Fig.\ref{fig:ArgonLFplotperiodic} demonstrates.  It should be pointed out, however, that in practice the $K$-matrix controlling channel interactions is never exactly a constant, and as a consequence, the periodicity will not hold accurately for the low members of any Rydberg series.

A number of positive features about these two types of traditional Lu-Fano plots are worth stressing, before we describe some situations where an alternative representation is advantageous, as the main point of this article:

{\it (i)} Both types of Lu-Fano plots can be implemented entirely from experimental data, even in the absence of any theoretical calculations.  For any bound state energy, assuming the two ionization threshold energies are known for a two-threshold system, the two effective quantum numbers can be computed at once and plotted in the form $-\nu_1$ (mod 1) versus either $\nu_2$ in the nonperiodic variant or versus $\nu_2$(mod 1) in the periodic Lu-Fano plot.  Recall that $-\nu_1$ (mod 1) = $\tau$ (mod 1) since any bound state obeys $\tau+\nu_1=n$ with $n$ an integer.   

{\it (ii)} Once the bound energy levels for a given symmetry are displayed on a Lu-Fano plot, it is apparent at a glance whether the channel interactions are quite strong or quite weak or somewhere in between.  This is because channel interaction strength shows up as avoided crossings of the branches of these plots, and if channel interactions are weak, the corresponding Lu-Fano plot has mostly vertical and horizontal branches with only very narrow avoided crossings between them.  

{\it (iii)} For systems with two ionization thresholds, the Lu-Fano plot shows how all bound energy levels, or in some applications autoionizing levels, are arranged on periodic curves whose pattern repeats each time $\nu_2 \rightarrow \nu_2+1$.  Moreover, those curves are usually fairly simple and smooth aside from branch changes, provided the energy splitting between the two threshold energies is not too small. 

Lu-Fano plots sometimes make other choices for the x-axis in the plots; the main point is that it is an energy axis in some sense.  Thus it is sometimes chosen to be the effective quantum number relative to the lower threshold, i.e. $\nu_1$, and sometimes when it is desirable to continue the curve above one or both ionization thresholds, it can simply be the energy in atomic units or cm$^{-1}$.  This flexibility is utilized in our design of the MLF presented below.

In the next section, we show how those periodic bound state curves on a traditional Lu-Fano plot get increasingly complex and rapidly varying, and therefore less informative when the threshold splitting gets tiny and the energy is far below threshold.  In that case, or when there are more than two ionization thresholds for the considered symmetry, an alternative representation of the $K$-matrix (denoted $\tilde{K}$) alongside a modified Lu-Fano plot (MLF) are demonstrated to be advantageous.


\section{Modified Lu-Fano Plot applied to Manganese}
The hyperfine states of the  $^6$P$^o_J$ and $^8$P$^o_J$ $np$ Rydberg series attached to the $^7$S$_3$ core of neutral manganese are an excellent showcase for where the traditional Lu-Fano curve begins to lose most of its value and how this alternative MLF representation provides advantages.
Because Mn has an open $3d$ subshell in its ionic core, there are occasional perturbers with entirely different character from the channel subspace attached to the $3d^5 (^6S) 4s$ ground state of the ion that we are considering throughout this analysis.  Perturbers for these Rydberg series are left out of the following calculations, since they occur at low energies and are largely absent from the Rydberg series with principal quantum numbers with $n>9$. Therefore, the energy dependent MQDT model used should be valid for n $\gtrsim$ 9, where the quantum defects remain largely unaffected by perturbers. 
To our knowledge, spectroscopic data of the hyperfine levels of these Rydberg series is not currently available in the literature, and thus the results below are predictions based on the nature of the low-lying state quantum defects, a frame transformation analysis, and MQDT.

Neutral manganese has 100\% natural abundance of its $^{55}$Mn isotope with nuclear spin of $I=5/2$, which splits its $^7$S$_3$ ionic ground state into 6 hyperfine thresholds with total ionic angular momentum quantum number $f_c=\frac{1}{2}\ ...\  \frac{11}{2}$. Using ionic hyperfine splitting constants from \cite{Blackwell-Whitehead_Toner_Hibbert_Webb_Ivarsson_2005}, we obtained hyperfine corrections to the $3d^5(^6S)4s\ ^7S_3$ ionic ground state of $^{55}$Mn from the following equation
\begin{equation}
    \Delta E_{\rm hfs} =\frac{1}{2}hA_JK+hB_J[K(K+1)-\frac{4}{3}I(I+1)J(J+1)]
\end{equation}
where $K=F(F+1)-I(I+1)-J(J+1)$. Using the experimental hyperfine parameters of $A_J=797.4(9.0)\ MHz$ and $B_J=0\ MHz$, the resulting hyperfine threshold splittings are \{-0.2660(30), -0.2261(25), -0.1596(18), -0.06650(75), 0.05320(60), 0.1995(23)\} cm$^{-1}$ for the total core hyperfine angular momentum values, $f_c$, 1/2, 3/2, 5/2, 7/2, 9/2, and 11/2 respectively, where 0 is set to the 2$f_c$+1 averaged energy.


The present discussion considers the neutral atomic $np$ Rydberg series of odd parity and example symmetries with total angular momenta $F=3$, $F=1$, and $F=0$, which have four, three, and two ionization thresholds respectively. The remaining symmetries for Mn $np$ Rydberg states are briefly described in Appendix B, along with an additional application of this MLF representation to hyperfine spectra of fermionic Yb.

\subsection{F=3}

Consider the $F=3$ symmetry of $^{55}$Mn, which has six Rydberg channels attached to four different hyperfine thresholds. We approximate the short range K-matrix to be diagonal in LSJ coupling with the short range coupling channels given by  
\begin{multline}
    \label{eq:LSJ3chan}
    \alpha =(^7Snp)^6P_{3/2},\ (^7Snp)^6P_{5/2},\ (^7Snp)^6P_{7/2},\\ (^7Snp)^8P_{5/2},\ (^7Snp)^8P_{7/2},\ (^7Snp)^8P_{9/2}.
\end{multline}
More concretely, the channel states $\alpha$ in the coupling scheme that is best suited to short distances or very low principal quantum numbers $n$ are given by $\ket{\alpha}=\ket{\{I [(S_c s_e)S, l_e]J\}\ F M_F  }$

Instead the coupling scheme appropriate at long range has the total core spin $S_c$ coupled with the core orbital angular momentum  $L_c$ to give the electronic angular momentum $J_c$ of the core. This is then coupled to the nuclear spin $I$ to give the total angular momentum $f_c$ of the ionic core.
The total hyperfine angular momentum of the core is then coupled with the total angular momentum $j_e$ of the Rydberg electron, where $\vec{j_e}=\vec{s_e}+\vec{\ell_e}$, to form a total $F$.
This coupling scheme can be written out as $\ket{i}=\ket{\{[I(S_cL_c)J_c]f_c,(s_el_e)j_e\}F M_F}$, 
and the six long range hyperfine coupling channels used are
\begin{multline}
\label{eq:fcj3chan}
i =(^7S)[\frac{3}{2}] \ p_{3/2}, (^7S)[\frac{5}{2}] \ p_{1/2}, (^7S)[\frac{5}{2}] \ p_{3/2},\\ (^7S)[\frac{7}{2}] \ p_{1/2}, (^7S)[\frac{7}{2}] \ p_{3/2}, (^7S)[\frac{9}{2}] \ p_{3/2}
\end{multline}
where the number in [ ] represents the total hyperfine angular momentum of the core, $f_c$.
The frame transformation matrix is now given by the following standard recoupling coefficient\cite{Robicheaux_Booth_Saffman_2018}:
\begin{figure}[t!]
    \centering
    \includegraphics[width=1\linewidth]{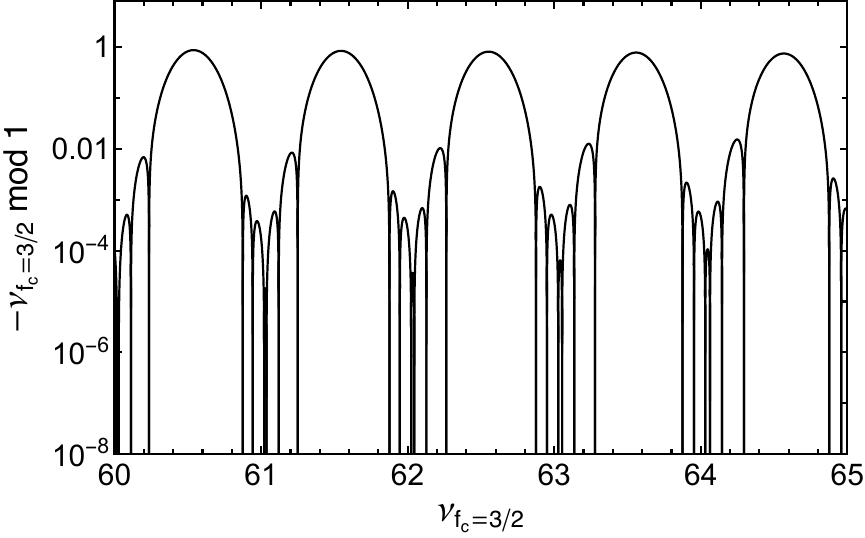}
    \caption{Plot of the determinantal bound state equation vs $\nu_{f_c=3/2}$ for the F=3 symmetry of $^{55}$Mn.}
    \label{fig:DetLog}
\end{figure}
\vspace{-0.5pt}
\begin{equation}
\begin{aligned}
\label{umat}
  \braket{i|\alpha}= U_{i \alpha}=\braket{(Ij_c)f_c\ j_e|I(SL)J}^{(F)}  =
    [S,L,j_c,j_e,f_c,J]\\
   (-1)^{I+j_c+j_e+J}
\begin{Bmatrix}
I & j_c & f_c \\
j_e & F & J
\end{Bmatrix}
\begin{Bmatrix}
s_c & s_e & S \\
l_c & l_e & L \\
j_c & j_e & J
\end{Bmatrix},
\end{aligned}
\end{equation}
where the notation [a,b,...] = $\sqrt{(2a+1)(2b+1)...}$.
 The energy dependent quantum defect parameters used can be found in Table. \ref{tab:MQDTparam} where the energy dependence of the quantum defect is defined as 
\begin{equation}
    \mu_\alpha(E)=\mu_\alpha^{(0)}+\mu_\alpha^{(1)}E +\mu_\alpha^{(2)}E^2+...
\end{equation}
where $E\equiv -1/2{\bar \nu}^2$ is the energy relative to the $(2f_c+1)$-weighted average of the hyperfine energy levels, and the reduced mass ratio has been factored into the expansion coefficients. These quantum defect parameters were obtained by fitting non-hyperfine resolved experimental energy levels from Refs. \cite{NIST_ASD,clear2025wavelengthsenergylevelsneutral}. 

As is summarized in various other review articles \cite{FanoJOSA, OrangeReview}, the hyperfine frame transformation consists here in the following steps:

{\it (a)}  Form the assumed-diagonal reaction matrix in the short-range eigenchannel representation:
\begin{equation}
    K^{\rm sr}_{\alpha,\alpha'}(\epsilon) = \tan{\pi \mu_\alpha(\epsilon)} \delta_{\alpha,\alpha'}
\end{equation}

{\it (b)} Transform $K$ into the long-range representation:
\begin{equation}
    K(E) = U K^{\rm sr}(E) U^T,
\end{equation}
where $T$ denotes the transpose.

{\it (c)}  Apply the MQDT equations, such as Eqs.\ref{eq:linearsystem1},\ref{eq:Rydbergform},\ref{eq:linearsystem2}, to determine bound state energies and their other derived properties such as transition oscillator strengths.   It is worth stressing at this point that the unitary transformation in step {\it (b)} would not accomplish anything by itself.  A crucial additional component of the MQDT  frame transformation theory is that the experimental threshold energies are now implemented in computing the effective quantum numbers $\nu_i$ that determine $\beta_i(E)$ in each channel.  This is the reason that the frame transformation is a powerful tool describing the interacting Rydberg channels, as opposed to being merely a unitary transformation of the basis.

\begin{figure}[t!]
    \centering
    \includegraphics[width=1\linewidth]{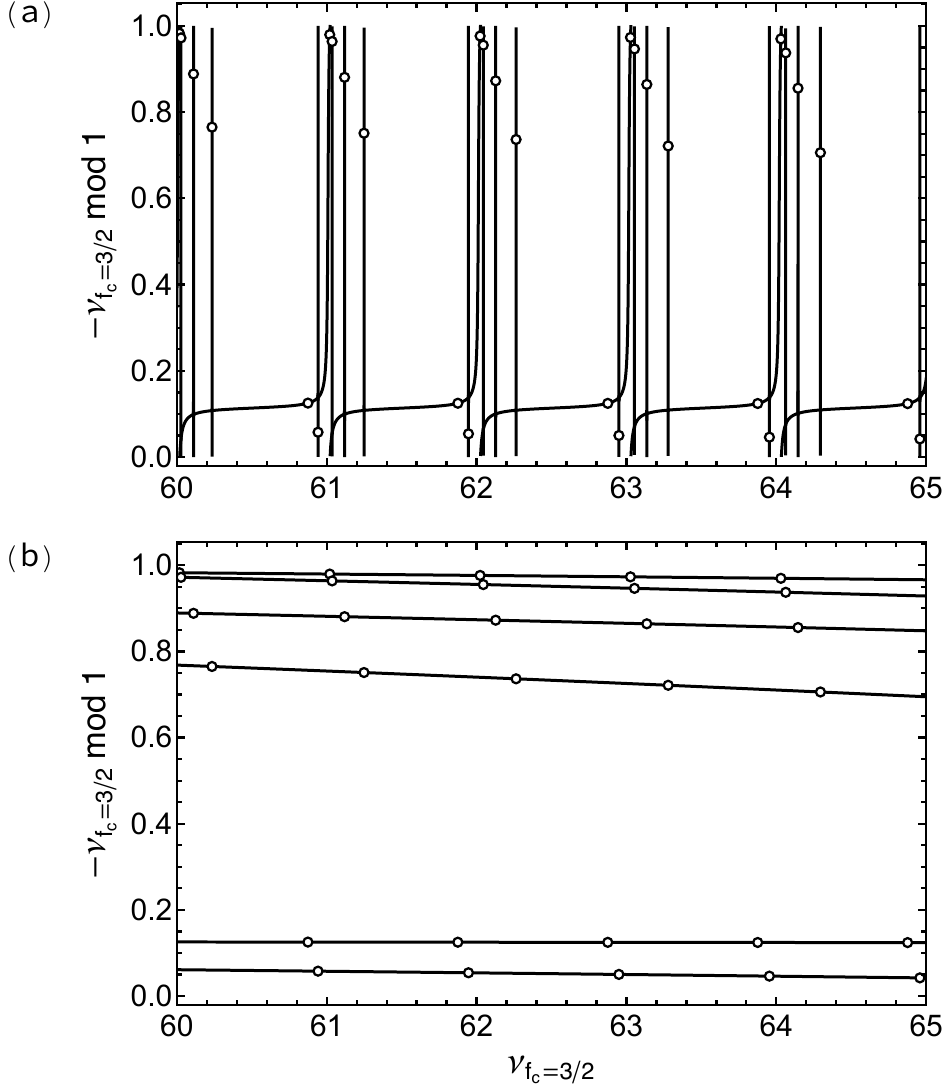}
    \caption{(a) Traditional Lu-Fano plot showing the eigenquantum defect -$\nu_{f_c=3/2}$ mod 1 vs $\nu_{f_c=3/2}$ for the F=3 symmetry of $^{55}$Mn. (b) Modified Lu-Fano curve showing the rotated eigenquantum defect -$\nu_{f_c=3/2}$ mod 1 vs $\nu_{f_c=3/2}$ for the F=3 symmetry of $^{55}$Mn. The open circles are calculated bound states. It can be seen how the modified Lu-Fano curve goes smoothly through the bound states in manner that is easy to interpret compared to the dense parallel trajectories of the traditional Lu-Fano curve.}
    \label{fig:LFF360t65Comp}
\end{figure}

Using {\it Method  (1)}, as described earlier in the text, would require us to solve for the roots of a highly oscillatory function in order to obtain bound state energies. As seen in Fig. \ref{fig:DetLog}, the four thresholds of the F=3 symmetry produce rapid oscillations with very tightly packed roots, which, as mentioned above, creates a nontrivial problem in accurately and systematically determining the roots over a large range.   
Implementation of {\it Method  (2)} gives a more traditional Lu-Fano curve which can be seen in Fig. \ref{fig:LFF360t65Comp}(a), which presents its own difficulties for this system.
Note that this plot chooses the effective quantum number $\nu_1$ of the lowest channel, $f_c=3/2$, on the x-axis, as opposed to the highest channel.  
This was done due to the the small splitting of the hyperfine thresholds causing differences of the effective quantum numbers between channels to be very small this far below threshold in addition to there being more than 2 thresholds. The number of thresholds increases the dimensionality required to show the periodic surface output of {\it Method (2)}. Therefore this projection of the 4-dimensional hypersurface was chosen for simplicity. While work has been done in making higher dimensional Lu-Fano curves easier to read \cite{Jin_Han_Gao_Zeng_Li_2017}, however, implementing such approaches are beyond the scope of this work. 


Fig.~\ref{fig:LFF360t65Comp}(a) showcases a limitation of the traditional Lu-Fano curve for systems with very small threshold splittings. In this case the curve is very rapidly changing as $\nu_{f_c=3/2}$ increases, and it is challenging to even group together states having similar character from one value of the principal quantum number to the next.  At the same time, the eye can pick out a far smoother set of curves that group together levels of similar character without oscillating so rapidly over each cycle of $\nu_{f_c=3/2}$.  Our explorations of many spectra have shown this to be a common situation when one is far below a set of ionization thresholds with extremely small splittings, as in Rydberg series converging to hyperfine-split thresholds.  This has motivated our determination of the smooth curves that the eye readily picks out, and which we denote the modified Lu-Fano (MLF) plot, defined next.


\begin{table*}
    \centering
    \begin{ruledtabular}
    {\footnotesize
    \begin{tabular}{ccccccc}
$\mu_\alpha$ &$(^7Snp)\ ^6P_{7/2}$ & $(^7Snp)\ ^6P_{5/2}$ & $(^7Snp)\ ^6P_{3/2}$ & $(^7Snp)\ ^8P_{5/2}$ & $(^7Snp)\ ^8P_{7/2}$ & $(^7Snp)\ ^8P_{9/2}$ \\
 $\mu_\alpha^{(0)}$ & 2.0410(7) & 2.0373(5) & 2.0373(4) & 2.1383(4) & 2.1366(3) & 2.1344(4)\\
 $\mu_\alpha^{(1)}(a.u.)^{-1}$ & 3.50(10) & 3.59(8) & 3.96(6) & -0.72(5) & -0.70(4) & -0.70(4)\\
 $\mu_\alpha^{(2)}(a.u.)^{-2}$ & 0 & 0 & 0 & 4(1) & 4(1) & 4(1) \\
    \end{tabular}}
    \end{ruledtabular}
    \caption{Energy dependent quantum defect fit parameters used for $^{55}$Mn, where the reduced mass ratio of 
    $1-\mu/m_e=9.98542\times10^{-6}$
    is factored into $\mu_\alpha^{(1)}$ and $\mu_\alpha^{(2)}$. These results were obtained from performing a fit on the fine structure resolved levels from NIST and other work \cite{NIST_ASD,clear2025wavelengthsenergylevelsneutral}.}
    \label{tab:MQDTparam}
\end{table*}

\textbf{Modified K-Matrix and Lu-Fano Curves}
Consider a rotation of the $(f_i,g_i)$ basis pair  in Eq. \ref{KmatrixSolutions} by the channel-dependent phase parameter equal to 
$\Delta\beta^{(i_0)}_i\equiv\beta_{i_0}-\beta_i$, where $i_0$ can be thought of as the pivot channel, or the channel in which you are rotating about.
Due to the aforementioned flexibility in choosing different options for the basis pair $(f,g)$ one can introduce a rotated version of $(f_i,g_i)$ which we denote as $(\tilde{f}_i^{({i_0})},\tilde{g}_i^{({i_0})})$ such that the linear combination that vanishes at infinity is formed instead using $\cos{\beta_{i_0}}$ and $\sin{\beta_{i_0}}$, i.e, using the negative energy phase parameter in any one chosen reference channel ${i_0}$ for {\it all} channels rather than the phase parameter in channel $i$, i.e. the same energy-normalized Whittaker function of Eq. \ref{Whittaker0} can then be recast as $W_i(r) = \tilde{f}_i^{({i_0})}(r) \cos{\beta_{i_0}}+\tilde{g}_i^{({i_0})}(r) \sin{\beta_{i_0}}$.  The linear combination that accomplishes this for channel $i$ is written in matrix form as:
\begin{equation}
\label{eq:rotationFG}
    \begin{pmatrix} \tilde{f}_i^{({i_0})}(r) \\ \tilde{g}_i^{({i_0})}(r) \end{pmatrix} = 
\begin{pmatrix} \cos{\Delta\beta_i^{({i_0})}} & -\sin{\Delta\beta_i^{({i_0})}} \\ \sin{\Delta\beta_i^{({i_0})}} & \cos{\Delta\beta_i^{({i_0})}} \end{pmatrix}
\begin{pmatrix} f_i(r) \\ g_i(r) \end{pmatrix} ,
\end{equation} 
It is useful for $i_0$ to be the lowest channel i.e. $i_0=1$, but other choices are possible. For generality the following equations will be written with a general $i_0$, but for all of the Mn spectra discussed in this study, $i_0$ is set to be the lowest channel threshold in the symmetry we are working in.  (For some Yb examples treated in Appendix C, $i_0$ will be chosen as the second channel, but the spectra would look very similar there with $i_0=1$.)  
\begin{figure}
    \centering
        \includegraphics[width=1\linewidth]{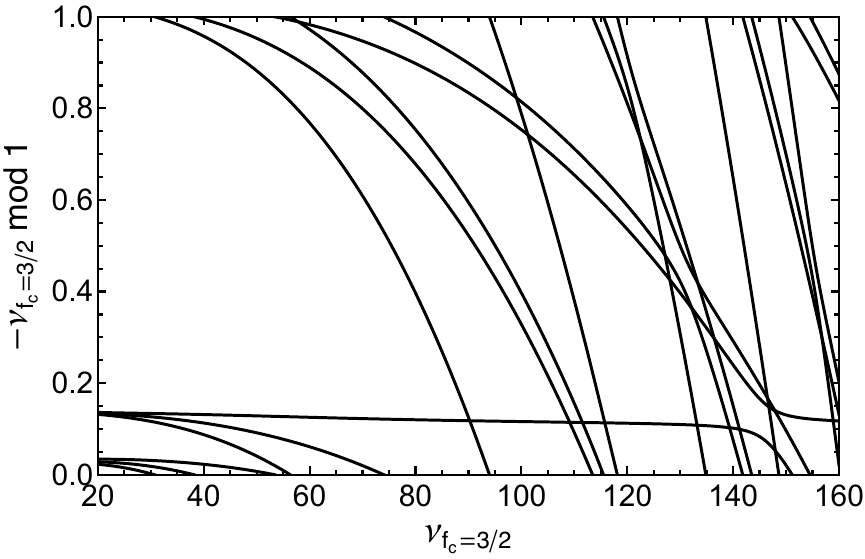}
    \caption{Plot of the modified Lu-Fano curve for the F=3 symmetry of $^{55}$Mn which shows the rotated eigenquantum defect -$\nu_{f_c=3/2}$ mod 1 vs $\nu_{f_c=3/2}$. This is just a larger range of Fig. \ref{fig:LFF360t65Comp}(b). Here the weakly interacting channels can be seen by the narrow crossings as opposed to the narrow lines as seen in Fig. \ref{fig:LFF360t65Comp}(a).}
    \label{fig:BLFF320t160}
\end{figure}
It can be seen how this transformation yields a new reaction matrix $\tilde{K}$ relatice to the rotated radial base pair, namely:
\begin{equation}
   \tilde{\psi}_{i'}=  {\cal A}\frac{1}{r} \sum_i \Phi_i(\Omega) (\tilde{f}_i^{({i_0})}(r) \delta_{ii'}-\tilde{g}_i^{({i_0})}(r) \tilde{K}_{i,i'}), \label{KmatrixSolutionsTilde}
\end{equation}
where 
\begin{equation}
    \tilde{K}=[\cos{\underline{\Delta\beta}}^{(i_0)}\ K-\sin{\underline{\Delta\beta}}^{(i_0)}][\cos{\underline{\Delta\beta}}^{(i_0)}+\sin{\underline{\Delta\beta}}^{(i_0)}\ K]^{-1}. \label{eq:Ktilde}
\end{equation}
Moreover, the modified set of linearly-independent $\tilde{K}$-matrix solutions $\tilde{\psi}_{i'}$ are related to the original set by
\begin{equation}
    \tilde{\psi}_{i'} =\sum_{i''} \psi_{i''} [\cos{\underline{\Delta\beta}^{(i_0)}}+ \sin{\underline{\Delta\beta}^{(i_0)}}K]^{-1}_{i'',i'}.
\end{equation}

\begin{figure}
    \centering
    \includegraphics[width=1\linewidth]{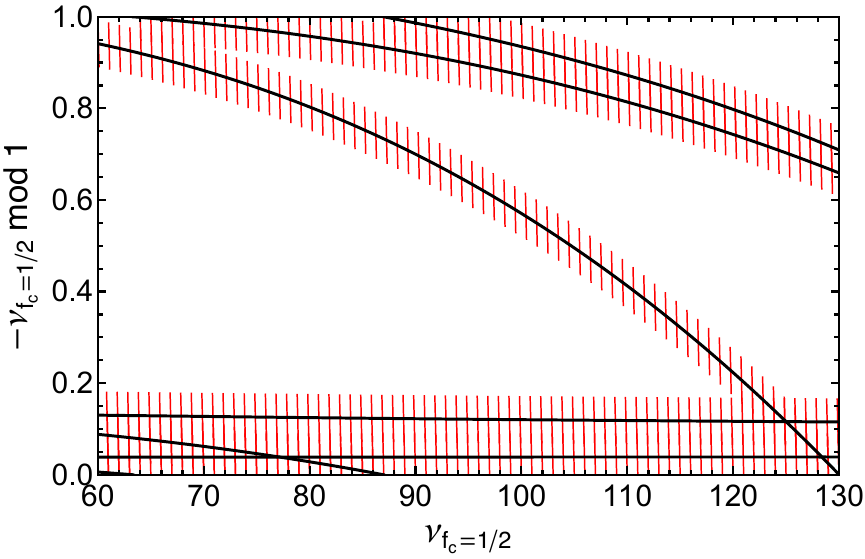}
    \caption{Plot of the modified Lu-Fano curve (black) for the F=1 symmetry of $^{55}$Mn with the quantization condition (red) corresponding to this modified representation.}
    \label{fig:BLFF1quantnob}
\end{figure}

It can be seen now what the rotation of basis functions, presented above, has accomplished. Namely, the linear combination of $\tilde{f}$ and $\tilde{g}$ that eliminates exponential growth is now {\it the same in every channel}, since: 
\begin{equation}
    \label{eq:fgtildlim}
    \lim_{r\rightarrow\infty}\tilde{f}_i(r)/\tilde{g}_i(r)\rightarrow-\tan(\Delta\beta^{(i_0)}_i+\beta_i)=-\tan\beta_{i_0},
\end{equation}
which gives a multichannel bound state condition of greater simplicity, i.e.:
\begin{equation}
    \label{eq:BLFbound}
    (\tan\beta_{i_0}\underline{\mathbb{I}}+\underline{\tilde{K}})\overrightarrow{B}=0.
\end{equation}
The arctangents of the eigenvalues of $\underline{\tilde{K}}(E)$, divided by $\pi$ are denoted the {\it rotated eigen quantum defects} $\tilde{\mu}_\alpha(E)$, and these smooth energy-dependent functions define what we call the modified Lu-Fano (MLF) curves.   The value of the transformation is made even clearer if we recast the determinantal eigenvalue equation, Eq. \ref{eq:BLFbound}, in the eigen representation, namely (assuming an integer value of the orbital angular momentum $l_{i_0}$):
\begin{equation}
    \label{eq:eigentilde}
    \prod_\alpha \sin{\pi(\nu_{i_0}+\tilde{\mu}_\alpha)}=0
\end{equation}
which is immediately solvable for the eigen energies, expressed in terms of their $\nu_{i_0}$  value, as $\nu_{i_0}(E^{(n,\alpha)})=n-\tilde{\mu}_\alpha(E^{(n)})$ with $n$ an integer.  Since the MLF curves $\tilde{\mu}_\alpha(E)$ are smooth, the bound state energies in the $\alpha$-th family are quickly found by iteration.  Alternatively, one can simply determine the roots graphically, as is illustrated in Fig.\ref{fig:BLFF1quantnob}, by finding the intersection of each MLF curve with one of the nearly-vertical red lines that correspond to the constraint $-\nu_{i_0} (\rm mod\ 1) = n-\nu_{i_o}$.

To recapitulate, the greater simplicity of the MLF curves  over the standard Lu-Fano plot curves derives from the difference in strategy.  As was mentioned before, the Lu-Fano style quantum defect curves are derived by first eliminating the exponential decay in all higher channels while temporarily treating those attached to the lowest threshold as though they are open.  When the system of interest has tiny threshold splittings and when the energy of interest is far below those closely-split thresholds, the traditional Lu-Fano curve is forced to oscillate as fast as the effective quantum number in the lowest channel, as is illustrated in Fig. \ref{fig:LFF360t65Comp}(a).  The contrast with the MLF curve, with slowly varying quantum defects $\tilde{\mu}_\alpha(E)$, shown in Fig. \ref{fig:LFF360t65Comp}(b) makes its advantages clear for this situation, which is common for Rydberg series converging to hyperfine-split ionization thresholds.  As a rule of thumb, we find that the Lu-Fano style plot works better once you are in a range where $d\nu_1 /d \nu_{\rm max} >>1$ , where $ \nu_{\rm max}$ is the effective quantum number referred to the highest threshold in the MQDT channel set.  On the other hand, the MLF style plot works better when $d\nu_1 /d \nu_{\rm max}$ is of order unity.  Thus the MLF is optimal when the energy range of interest is far below threshold, in units of the maximum threshold splitting.

As an illustration of the resulting simplicity of the MLF strategy, the rotated K-matrix $\tilde{K}$ for this system has been constructed and diagonalized  as a function of $\nu_{f_c=3/2}$, which yields the MLF curves shown in Fig. \ref{fig:BLFF320t160}, which covers several decades of $\nu_{f_c=3/2}$, and the narrower region displayed in Fig. \ref{fig:LFF360t65Comp}(b).
The curves in Fig. \ref{fig:LFF360t65Comp}(b) run smoothly through the bound state energies, with the behavior one has grown to expect from a traditional Lu-Fano plot for any system with larger threshold splittings of order 0.05 eV or larger. Comparing this to the curve in Fig. \ref{fig:LFF360t65Comp}(a) shows the increase in visual readability the MLF curve of Fig. \ref{fig:LFF360t65Comp}(b) presents. 
This smoothness arises due to the rotation by $\Delta\beta_i^{(i_0)}$ effectively suppressing the rapid oscillations brought about by the $\tan\beta$ from Eq. \ref{eq:linearsystem1}. When the effective quantum numbers $\nu_i$ are expressed as functions of the effective quantum number of the lowest channel $\nu_1$, channels with small threshold splittings vary at nearly the same rate as $\nu_1$, causing rapid oscillations in the traditional Lu-Fano curve. For closely split thresholds the $\Delta\beta_i^{(i_0)}$ is a slowly varying function of energy. 
Careful readers may notice how the introduction of perturbing states attached to even higher thresholds outside of the chosen MQDT channel set would cause the $\Delta\beta_i^{(i_0)}$ to rapidly vary with energy, thus disrupting the smoothness of the curve. This does not occur in the Mn spectra we analyze here, but when it does happen it poses no difficulty and an example is shown in Appendix C where the MLF curve is applied to $^{171}$Yb and compared to experiment\cite{Peper_Li_Knapp_Bileska_Ma_Liu_Peng_Zhang_Horvath_Burgers_2025}.


\subsection{F=1}
Another symmetry which illustrates the usefulness of the MLF curve is the F=1 symmetry of $^{55}$Mn. This symmetry has five channels converging to three hyperfine thresholds, with the LSJ coupled short range channels and long range hyperfine coupled channels used being:  

\begin{multline}
    \label{eq:LSJ1chan}
    \alpha =(^7Snp)^6P_{3/2},\ (^7Snp)^6P_{5/2},\ (^7Snp)^6P_{7/2},\\ (^7Snp)^8P_{5/2},\ (^7Snp)^8P_{7/2}
\end{multline}
\begin{multline}
   \label{eq:fcj1chan}
    i =(^7S)[\frac{1}{2}] \ p_{1/2}, (^7S)[\frac{1}{2}] \ p_{3/2}, (^7S)[\frac{3}{2}] \ p_{1/2},\\ (^7S)[\frac{3}{2}] \ p_{3/2}, (^7S)[\frac{5}{2}] \ p_{3/2}
\end{multline}
respectively, where the energy dependent MQDT parameters used can be found in Table \ref{tab:MQDTparam}.

Diagonalizing the rotated $K$-matrix $\tilde{K}$ produces the black curve shown in Fig. \ref{fig:BLFF1quantnob}, where the red lines indicate the new quantization condition discussed above, with intersections of these curves correspond to the bound state energies. 
Similar to Fig. \ref{fig:BLFF320t160}, the MLF representation makes it straightforward to identify regions of near degeneracy and to visualize the strength of channel interactions.

\subsection{F=0}
The simplest symmetry of $^{55}$Mn, with more than one channel, F=0 is a two channel problem converging to two hyperfine thresholds. The LSJ coupled short range channels and long range hyperfine coupled channels used are 
\begin{equation}
    \label{eq:LSJ0chan}
    \alpha =\ (^7Snp)\ ^6P_{5/2},\ (^7Snp)\ ^8P_{5/2} 
\end{equation}
\begin{equation}
\label{eq:fcj0chan}
i =\ (^7S)[\frac{1}{2}] \ p_{1/2},\ (^7S)[\frac{3}{2}] \ p_{3/2},
\end{equation}
respectively. The energy dependent MQDT parameters are found in Table \ref{tab:MQDTparam} and the 2x2 frame transformation for this symmetry can be found in Table \ref{tab:MQDT0param}.

\begin{figure}[!b]
    \centering
    \includegraphics[width=1\linewidth]{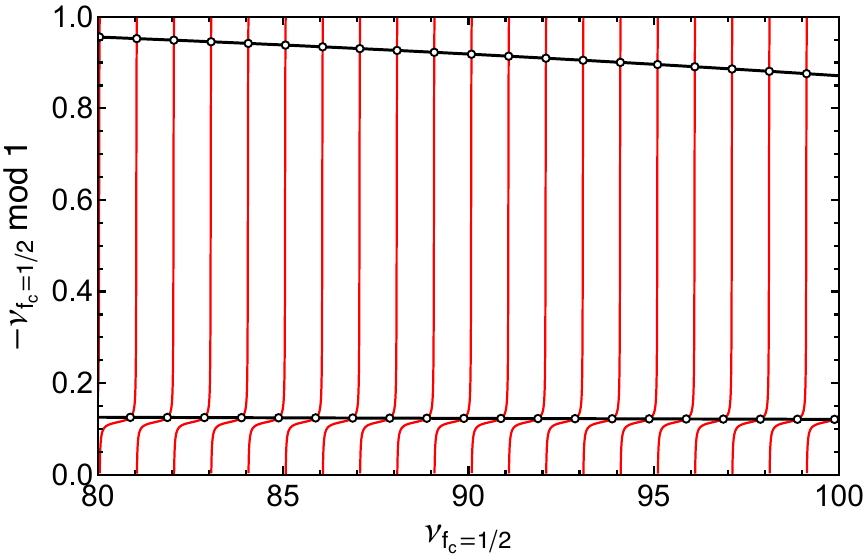}
    \caption{Plot of the both the modified Lu-Fano curve (black) and the traditional Lu-Fano curve (red) for the F=0 symmetry of $^{55}$Mn. The open circles are calculated bound state energies. }
    \label{fig:BLFF0LFComp}
\end{figure}

Having two thresholds, the F=0 symmetry of $^{55}$Mn fits the criteria of systems in which the traditional Lu-Fano curve is most useful. While the bound state intersections of the Lu-Fano curve are used for systems with hyperfine splittings \cite{Peper_Li_Knapp_Bileska_Ma_Liu_Peng_Zhang_Horvath_Burgers_2025} the curve itself is of little use due to reasons that were discussed above. However, the MLF curve is of great use as seen in Fig. \ref{fig:BLFF0LFComp}. Fig. \ref{fig:BLFF0LFComp} shows the rotated eigenquantum defects for the F=0 symmetry with the traditional Lu-Fano curve visible in the same figure, which again shows the improved visual clarity the MLF curve presents over the traditional Lu-Fano curve, for systems like this one with closely split thresholds.

\begin{table}[b]
    \centering
    \begin{tabular}{cccc}
    \toprule
 & &$\alpha$ = 1 & 2 \\
$U_{i\alpha}$ & $i$ = 1 & $-\sqrt{\frac{5}{21}}$ & $\frac{4}{\sqrt{21}}$  \\
 & 2 & $\frac{4}{\sqrt{21}}$ & $\sqrt{\frac{5}{21}}$ \\
 \toprule
    \end{tabular}
    \caption{F=0 frame transformation matrix }
    \label{tab:MQDT0param}
\end{table}

\section{Conclusions and Outlook}
Graphical representations of atomic and molecular Rydberg spectra are a useful tool for theory and experimental work alike. They bring out regularities in the spectra and help to classify and connect energy eigenstates of similar character.  Beyond their interpretive value, they can also help to catch errors, as in the energy level displayed incorrectly in the NIST tables \cite{NIST_ASD} that we discovered in the course of creating Figs. \ref{fig:ArgonLFplot} and \ref{fig:ArgonLFplotperiodic} for atomic argon.

The rotation of the Coulomb $(f_i,g_i)$ basis functions by a $\Delta\beta^{(i_0)}_i$ in the wavefunction of the Rydberg electron, seen in Eq. \ref{eq:rotationFG}, allows for an alternate representation of the K-matrix as a rotated K-matrix $\underline{\tilde{K}}$ seen in Eq. \ref{eq:Ktilde}.
This alternate MLF representation leads to a simplified quantization condition and a MLF curve, differing from the Lu-Fano plot curves, which smoothly intersects all bound state energies in a manner that is visually intuitive for systems with closely split thresholds.

Applications to the $^6$P$^o_J$ and $^8$P$^o_J$ Rydberg series of $^{55}$Mn that converge to many hyperfine-split ionization thresholds have shown the usefulness of the MLF plot and corresponding curves.  This representation should be useful to experimentalists as the MLF curve presents a highly readable plot of the channel interaction strengths, and provides an aid to interpreting experimental data and identifying errors. Further examples of this strategy can be found in Appendix C where we utilize this representation to compare to accurate Rydberg energy level data for $^{171}$Yb\cite{Peper_Li_Knapp_Bileska_Ma_Liu_Peng_Zhang_Horvath_Burgers_2025,GallagherYb}.

\acknowledgements
This work is supported in part by the U.S. Department of Energy, Office
of Science, Basic Energy Sciences, under Award No. DESC0010545.

\appendix




\section{Normalization Integral}
Following procedure laid out in Refs.\cite{Greene_Fano_Strinati_1979,Greene_Rau_Fano_1982,Greene_Rau_Fano_1984,Lee_Lu_1973,Seaton_1966,Seaton_RPP_1983}, 
the normalization of $\Psi^{(n)}=\sum_{i'}\psi_{i'}A^{(n)}_{i'}$, with $\psi_{i'}$ being given by 
\begin{multline}
\psi_{i'}=\sum_{i}\phi_i(\tilde{f_i}[\cos{\Delta\beta^{(i_0)}_i}\delta_{ii'}+\sin{\Delta\beta^{(i_0)}_i}K_{ii'}]\\-\tilde{g_i}[\cos{\Delta\beta^{(i_0)}_i}K_{ii'}-\sin{\Delta\beta^{(i_0)}_i}\delta_{ii'}]),
\label{eq:psiCStilde}
\end{multline}
in the modified representation is carried out below. For comparison the normalization of the traditional K-matrix form of Eq. \ref{KmatrixSolutions} is given after as well. 
First, rewrite Eq. \ref{eq:psiCStilde} with 
\begin{eqnarray*}
    \tilde{C}_{ii'}\equiv\cos{\Delta\beta^{(i_0)}_i}\delta_{ii'}+\sin{\Delta\beta^{(i_0)}_i}K_{ii'}\\
    \tilde{S}_{ii'}\equiv\cos{\Delta\beta^{(i_0)}_i}K_{ii'}-\sin{\Delta\beta^{(i_0)}_i}\delta_{ii'}
\end{eqnarray*}
such that Eq. \ref{eq:psiCStilde} takes the form $\psi_{i'}=\sum_{i}\phi_i(\tilde{f_i}\tilde{C}_{ii'}-\tilde{g_i}\tilde{S}_{ii'}),$ and $\underline{\tilde{K}}\equiv\underline{\tilde{S}}\ \underline{\tilde{C}}^{-1}$. From here it can be seen how \cite{Greene_Fano_Strinati_1979} 
\begin{multline}
    \int\Psi_{E'}\Psi^{(n)}_E d\tau=\\
    \frac{1}{\pi}\sum_{i\,i'i''}\frac{d}{dE}[\sin\beta_{i_0}\tilde{C}_{ii'}+\cos\beta_{i_0}\tilde{S}_{ii'}]|_{E^{(n)}}\\
[\sin\beta_{i_0}\tilde{S}_{ii'}-\cos\beta_{i_0}\tilde{C}_{ii'}]A^{(n)}_{i'}A^{(n)}_{i''}
\end{multline}
where $A$ is the solution to the linear system of equations: $(\sin\beta_{i_0}\ \underline{\tilde{C}}+\cos\beta_{i_0}\ \underline{\tilde{S}})\overrightarrow{A}^{(n)}=0$. This system can be expressed in terms of the rotated K-matrix $\underline{\tilde{K}}$ through the insertion of identity, which gives $(\tan\beta_{i_0}\underline{\mathbb{I}}+\underline{\tilde{K}})\overrightarrow{B}^{(n)}=0$ where $\overrightarrow{B}^{(n)}=\underline{\tilde{C}}\ \overrightarrow{A}^{(n)}$ and are the eigenvectors of $\underline{\tilde{K}}$. We can now write the bound state wave function as \cite{OrangeReview} 
\begin{equation}
    \Psi^{(n)}=\sum_i\frac{1}{r}\Phi_i(\omega)W_i(r,\nu_i,\ell_i)Z^{(n)}_i
\end{equation}
where
\begin{equation}
    Z^{(n)}_i=\frac{\nu^{3/2}_i}{\cos\beta_{i_0}}B_i^{(n)}/\tilde{N}_n.
\end{equation}
These $Z^{(n)}_i$ coefficients satisfy the normalization condition 
\begin{equation}
    \sum_{i\,i'}Z_i^{(n)}\zeta_{ii'}Z_{i'}^{(n)}=1
    \label{eq:normcond}
\end{equation}
where
\begin{equation}
    \zeta_{ii'}=\delta_{ii'}+\cos\beta_{i}\cos\beta_{i_0}\bigg(\frac{2}{\pi\nu_i^3}\bigg)^{1/2}\tilde{D}_{ii'}\bigg(\frac{2}{\pi\nu_{i'}^3}\bigg)^{1/2}
\end{equation}
with
\begin{equation*}
    \tilde{D}_{ii'}=\sum_{p}\frac{dK_{ip}}{dE}\bigg|_{E^{(n)}}\tilde{C}^{-1}_{pi'}.
\end{equation*}

Similar to what was done above, for $\psi_{i'}$ in the form of Eq. \ref{KmatrixSolutions} the normalization integral takes the form of the following:
\begin{multline}
    \int\Psi_{E'}\Psi^{(n)}_E d\tau=\\\frac{1}{\pi}\sum_{i\,i'i''}\frac{d}{dE}[\sin\beta_i\delta_{ii'}+\cos\beta_iK_{ii'}]|_{E^{(n)}} \\ [\sin\beta_iK_{ii'}-\cos\beta_i\delta_{ii'}]A^{(n)}_{i'}A^{(n)}_{i''},
\end{multline}
where $A$ is the solution to the linear system given in Eq. \ref{eq:linearsystem1}. This gives the familiar form for the $Z^{(n)}_i$ coefficients of 
\begin{equation}
    Z^{(n)}_i=\frac{\nu^{3/2}_i}{\cos\beta_i}A_i^{(n)}/N_n
\end{equation}
which satisfy the same normalization condition seen in Eq. \ref{eq:normcond} for
\begin{equation}
    \zeta_{ii'}=\delta_{ii'}+\cos\beta_i\bigg(\frac{2}{\pi\nu_i^3}\bigg)^{1/2}\frac{dK_{ii'}}{dE}\bigg|_{E^{(n)}}\bigg(\frac{2}{\pi\nu_{i'}^3}\bigg)^{1/2}\cos\beta_{i'}.
\end{equation}


\section{Remaining Symmetries of $^{55}$Mn}
$^{55}$Mn has 8 different symmetries of F, where we only covered 3 of the 8 in the main text. For completion we will briefly cover 4 of the remaining 5 symmetries, as the F=7 symmetry is a trivial single channel problem existing only in the $(^7S)[\frac{11}{2}]\ p_{3/2}$ channel.

\subsection{F=2 \& F=4}
The F=2 and F=4 symmetries have six channels going to four hyperfine thresholds, similar to F=3. The short range and long range channels, for F=2, used in the calculation given as the following
\begin{multline}
    \label{eq:LSJ2chan}
    \alpha =(^7Snp)^6P_{3/2},\ (^7Snp)^6P_{5/2},\ (^7Snp)^6P_{7/2},\\ (^7Snp)^8P_{5/2},\ (^7Snp)^8P_{7/2},\ (^7Snp)^8P_{9/2}
\end{multline}
\begin{multline}
\label{eq:fcj2chan}
i =(^7S)[\frac{1}{2}] \ p_{3/2}, (^7S)[\frac{3}{2}] \ p_{1/2}, (^7S)[\frac{3}{2}] \ p_{3/2},\\ (^7S)[\frac{5}{2}] \ p_{1/2}, (^7S)[\frac{5}{2}] \ p_{3/2}, (^7S)[\frac{7}{2}] \ p_{3/2}.
\end{multline}
Where the short range channels for F=4 are the same as F=2 and the $f_c$ of the long range channels are incremented by 1. Using this to build the rotated K-matrix $\underline{\tilde{K}}$ for these two systems allows us to obtain the rotated eigenquantum defects in the form of a MLF curve as seen in Fig. \ref{fig:BLFF2456}(a) for F=2 and Fig. \ref{fig:BLFF2456}(b) for F=4.



\subsection{F=5}
\begin{figure}[t]
    \centering
    \includegraphics[width=1\linewidth]{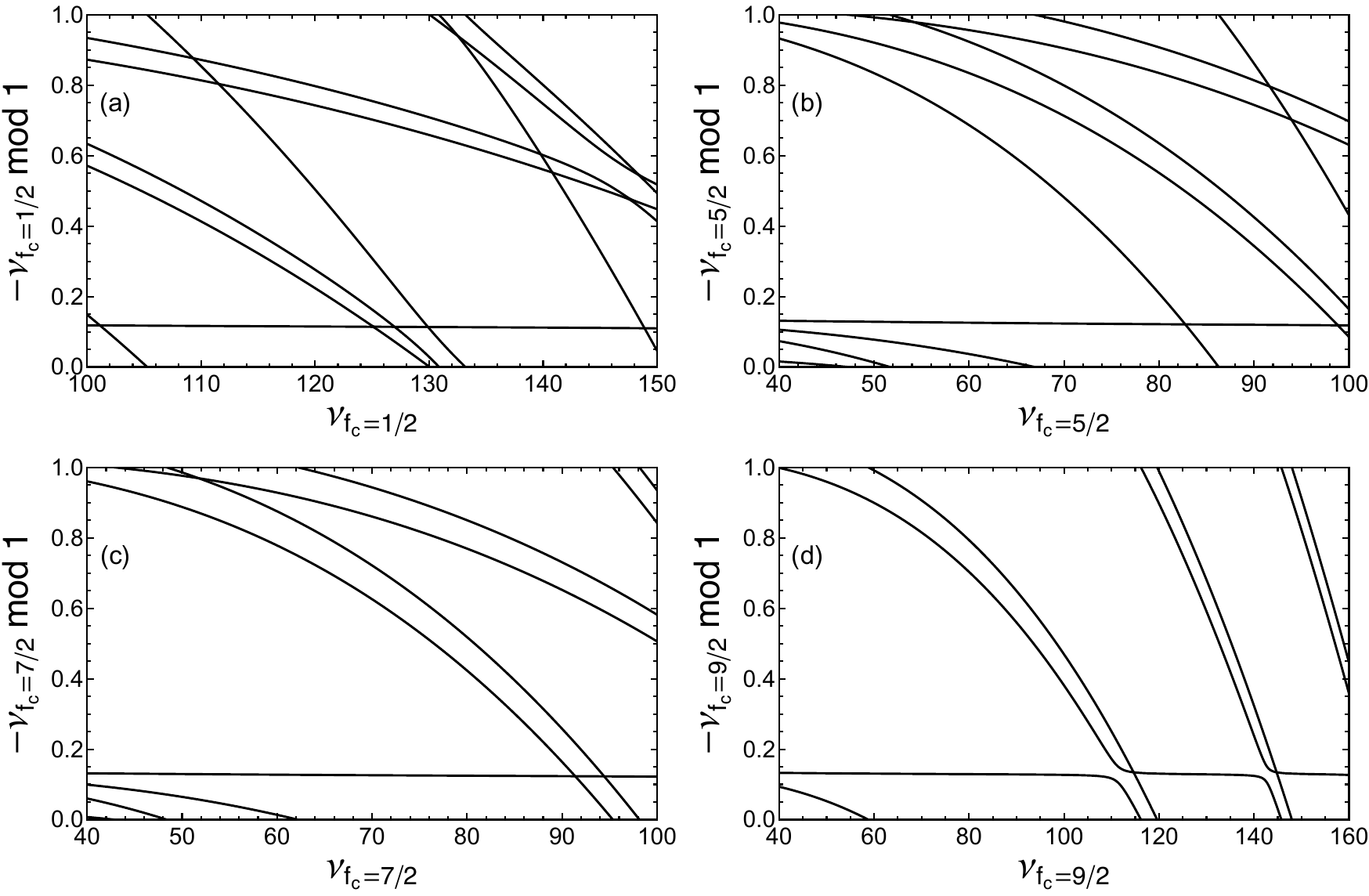}    
    \caption{MLF curves for the following symmetries (a) F=2 (b) F=4 (c) F=5 (d) F=6 }
    \label{fig:BLFF2456}
\end{figure}

The F=5 symmetry has five channels going to 3 hyperfine thresholds with short and long range channels given by
\begin{multline}
    \label{eq:LSJ5chan}
    \alpha =(^7Snp)^6P_{5/2},\ (^7Snp)^6P_{7/2},\ (^7Snp)^8P_{5/2},\\ (^7Snp)^8P_{7/2},\ (^7Snp)^8P_{9/2}
\end{multline}
\begin{multline}
\label{eq:fcj5chan}
i =(^7S)[\frac{7}{2}] \ p_{3/2}, (^7S)[\frac{9}{2}] \ p_{1/2}, (^7S)[\frac{9}{2}] \ p_{3/2},\\ (^7S)[\frac{11}{2}] \ p_{1/2}, (^7S)[\frac{11}{2}] \ p_{3/2}.
\end{multline}
Fig. \ref{fig:BLFF2456}(c) shows the MLF curve for this symmetry.

\subsection{F=6}
The F=6 symmetry has three channels going to 2 hyperfine thresholds with short and long range channels given by
\begin{multline}
    \label{eq:LSJ5chan}
    \alpha =(^7Snp)^6P_{7/2},\ (^7Snp)^8P_{7/2},\ (^7Snp)^8P_{9/2}
\end{multline}
\begin{multline}
\label{eq:fcj5chan}
i =(^7S)[\frac{9}{2}] \ p_{3/2}, (^7S)[\frac{11}{2}] \ p_{1/2}, (^7S)[\frac{11}{2}] \ p_{3/2}.
\end{multline}
Fig. \ref{fig:BLFF2456}(d) shows the MLF curve for this symmetry.



\begin{figure}[!b]
    \centering
    \includegraphics[width=1\linewidth]{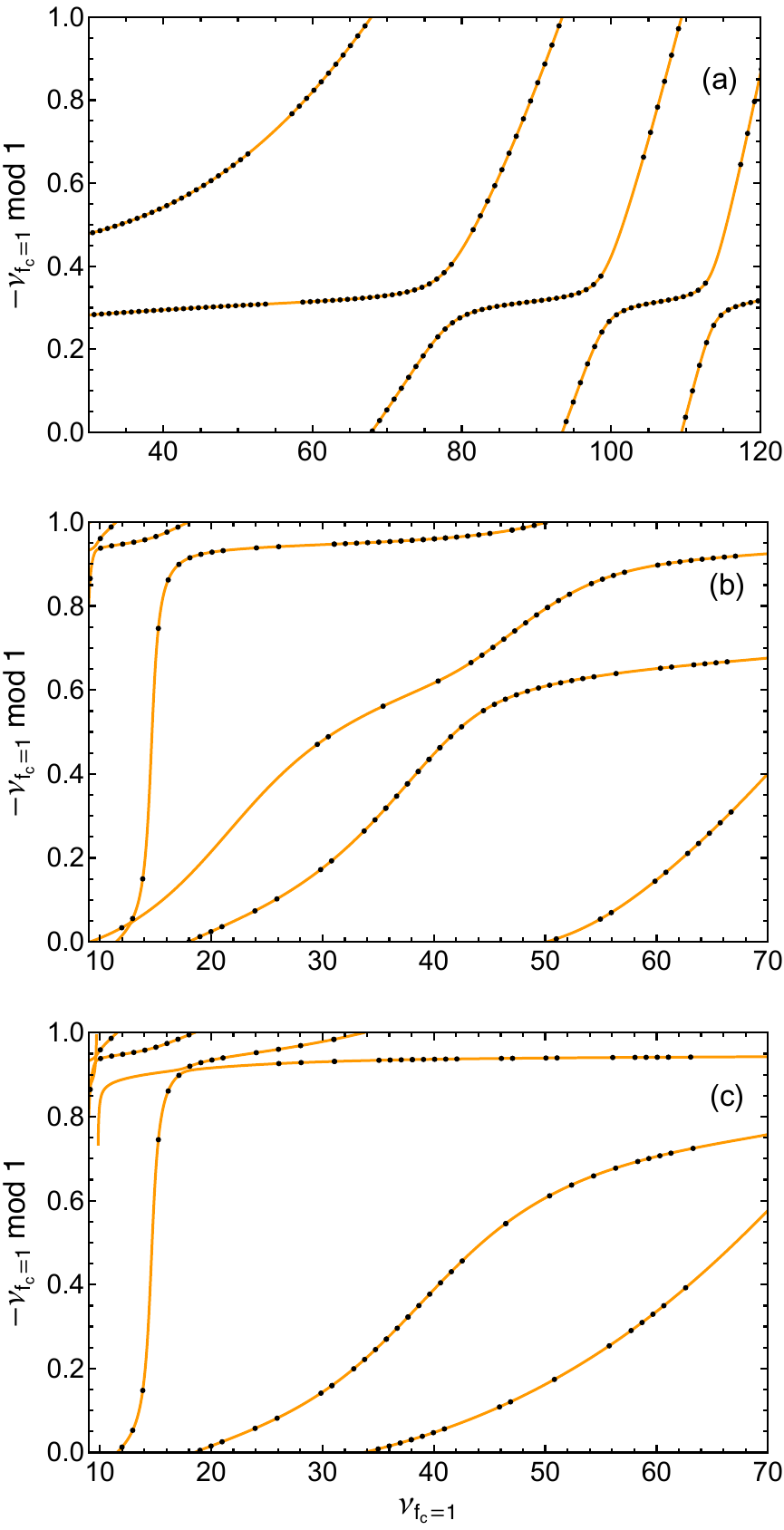}
    \caption{Plots of the MLF curve (orange) along side experimental data (black points) for $^{171}$Yb for the symmetries (a) $\ket{\nu ,L=0, F=1/2}$, (b) $\ket{\nu, L=1, F=1/2}$, and (c) $\ket{\nu, L=1, F=3/2}$.}
    \label{fig:Ybplots}
\end{figure}
\section{$^{171}$Yb}
$^{171}$Yb is an atom with a two level closely split hyperfine system that has been studied profusely in the effort to build efficient two-qubit gates. As a result many studies have been published containing both experimental and theoretical data, making it perfect for another example application of the Modified Lu-Fano representation. Using the detailed data from the supplement of the article by Peper {it et al.}\cite{Peper_Li_Knapp_Bileska_Ma_Liu_Peng_Zhang_Horvath_Burgers_2025} we show several of the resulting MLF-style plot curves in comparison with the experimental energies. 

The channels used in these MQDT models include a small number of perturbers associated with much higher thresholds than the closely split hyperfine thresholds, characterized in Ref.\cite{Peper_Li_Knapp_Bileska_Ma_Liu_Peng_Zhang_Horvath_Burgers_2025} through a fitting matrix that mixes these perturbers into the hyperfine channels of interest. This presents an important step that must be taken in order to properly produce a MLF curve for situations such as this. 
There are two ways in which this can be dealt with.
Before ``rotating" the K-matrix to form $\tilde{K}$ one should first perform the channel elimination in the strongly closed channels going to the much higher thresholds, in order to obtain a $\underline{K}^{\rm red}$ as seen in Eq. \ref{eq:channelelim}. Where the ``open", near(n), channels will include only the channels attached to the closely-split  thresholds
and the ``closed", far(f), channels will include only the channels attached to much higher thresholds: 

\begin{equation}
    K^{\rm red}=K_{\rm nn}-K_{\rm nf}(K_{\rm ff}+\tan\pi\nu_f)^{-1}K_{\rm fn}.
\end{equation}
\begin{figure*}[t!]
    \centering
    \includegraphics[width=1\linewidth]{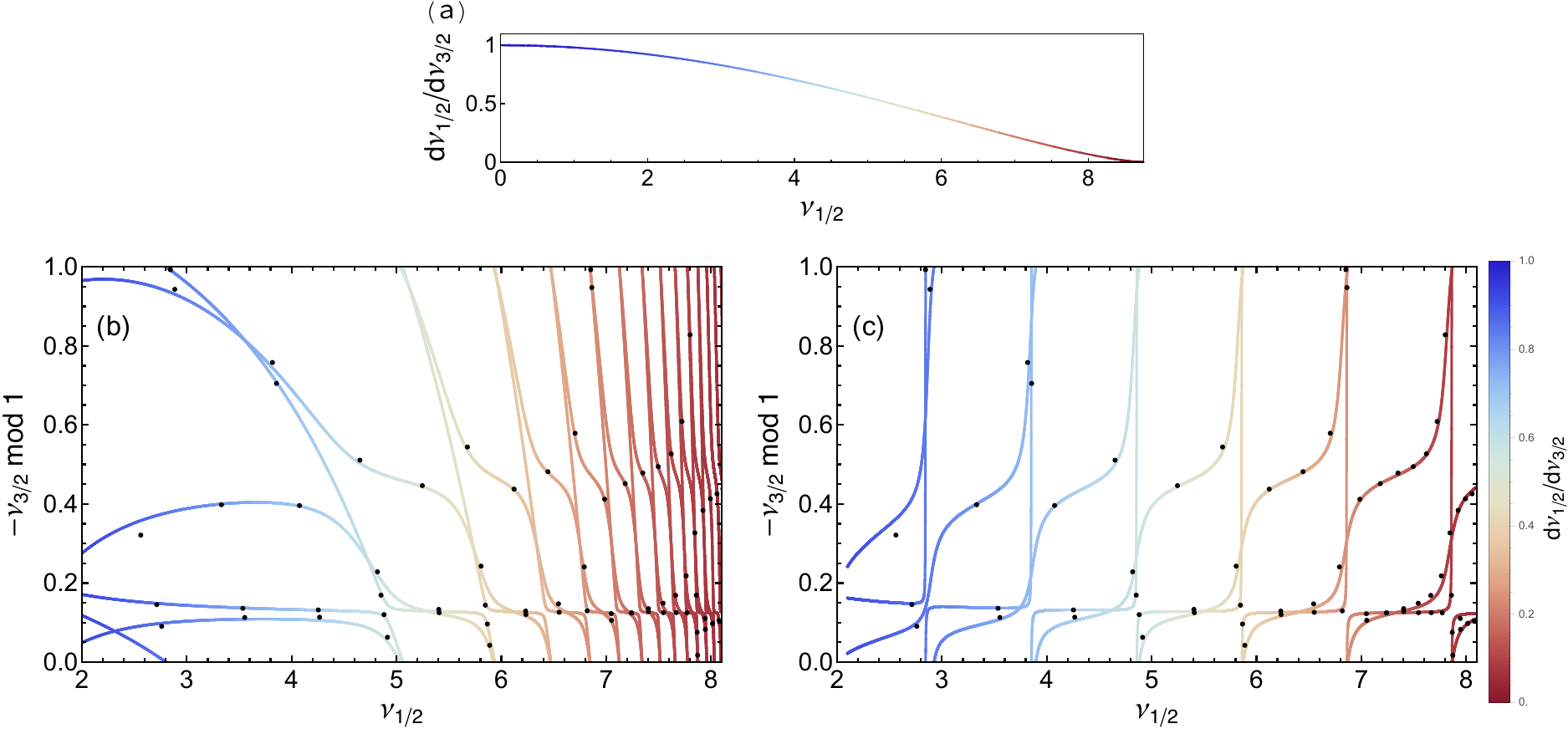}
    \caption{ (a) $d\nu_{\rm max}/d\nu_{1}$ for the $J^\pi=1^-$ symmetry for Ar, the threshold splitting for this system is $E_{1/2}-E_{3/2}=1431.5831$cm$^{-1}$ \cite{NIST_ASD}. The color gradient seen throughout this figure is relative to the magnitude of $d\nu_{\rm max}/d\nu_{1}$. (b) MLF plot, for the system described above, colored in accordance with $d\nu_{\rm max}/d\nu_{1}$ for this system. When $d\nu_{\rm max}/d\nu_{1}\approx1$ it can be seen how the MLF curve provides a useful depiction of the Rydberg spectra. The black points are experimental bound states taken from NIST \cite{NIST_ASD}. (c) Same as Fig. \ref{fig:ArgonLFplot} except with a color gradient showing for the regions where $d\nu_{\rm max}/d\nu_{1}\approx0$ the Lu-Fano plot better depicts the Rydberg spectra over the MLF plot.}
    \label{fig:ArgonMLF}
\end{figure*}
From here $\tilde{\underline{K}}$ will be obtained using the $K^{\rm red}$ instead of the entire $K$-matrix. Alternatively, one could obtain $\tilde{\underline{K}}$ from the whole $K$-matrix, including the perturbing channels, and perform channel elimination for the same strongly-closed channels as described above with the difference being $\tan\pi\nu_j$ in the parenthesis as seen below
\begin{equation}
    \tilde{K}^{\rm red}=\tilde{K}_{\rm nn}-\tilde{K}_{\rm nf}(\tilde{K}_{\rm ff}+\tan\pi\nu_{i_0})^{-1}\tilde{K}_{\rm fn}.
\end{equation}
Both methods result in the same matrix so one can choose the strategy based on convenience. 
Fig. \ref{fig:Ybplots} contains the resulting MLF curves for the $nS$ $F=1/2$, $nP$ $F=1/2$, and $nP$ $F=3/2$ symmetries for $^{171}$Yb. These examples show the MLF curve applied to a system where fitted MQDT parameters allow a detailed comparison with experimental data.


\section{Validity of MLF Plot}
\vspace{-3pt}
Throughout the paper we have made much mention of closely split thresholds for the applicability of the MLF plot.
However, the question of how closely split the thresholds have to be remains. To answer this question, consider the quantity $d\nu_{max}/d\nu_{1}$, where the general derivative of $\nu_i$ and $\nu_j$ is given by
\begin{equation}
    \frac{d\nu_{i}}{d\nu_{j}}=\bigg(\frac{\nu_i}{\nu_{j}}\bigg)^{3},
\end{equation}
with $\nu_i$ and $\nu_j$ the effective quantum numbers relative to thresholds $E_i$ and $E_j$, respectively.
As is mentioned in the main text, the MLF plot is most effective when $\Delta\beta$ varies slowly with energy, and using  $d\nu_{max}/d\nu_{1}$ allows an immediate assessment of how rapidly $\Delta\beta$ varies at a given energy. In principle, one would want to look at $d\nu_{i}/d\nu_{i_0}$ in each channel to ensure the full vector of $\Delta\beta_i^{(i_0)}$ is a slowly varying function of energy, but in practice the information is mostly contained in this derivative, with $i$ the highest threshold in the MQDT channel set.

\vspace{-1pt}
Applying this to the $J^\pi = 1^-$ symmetry of Ar \cite{Lee_Lu_1973}, as seen in Fig. \ref{fig:ArgonMLF}, illustrates the regions in which the MLF representation is effective, those where the traditional Lu–Fano plot is more appropriate, and the transition region between them.
Observe that in Fig. \ref{fig:ArgonMLF}, it is evident  that when $d\nu_{\rm max}/d\nu_{1} \approx 1$, the MLF curve passes smoothly through the bound-state levels,
but when $d\nu_{\rm max}/d\nu_1$ decreases below 0.5 the MLF curve begins to rapidly oscillate. 
It is in this latter region where $d\nu_{\rm max}/d\nu_1$ approaches 0, corresponding to energies approaching the threshold, where the 
traditional Lu-Fano plot provides a more useful depiction of the Rydberg spectra. Performing this type of analysis for other non-hyperfine split problems may reveal that the MLF curve can provide a useful graphical representation of Rydberg spectra. For example, we believe that the MLF plot would be helpful for such as the $J^\pi=0^+$ autoionizing levels of Ca between the $4s$ and $3d_{3/2}$ thresholds as seen in \cite{OrangeReview,Aymar_Telmini_1991}, $J^\pi=0^+$ autoionizing levels of Sr between the $5s$ and $4d_{3/2}$ thresholds \cite{OrangeReview,Kompitsas_Goutis_Aymar_Camus_1991}, and the $4fnf$ $J=6$ autoionizing levels of Ba \cite{OrangeReview,Luc-Koenig1992-ld} to name a few.

\end{document}